\documentclass[submission,copyright,creativecommons]{eptcs}
\providecommand{\event}{FORECAST 2016} 
\usepackage{enumerate}
\usepackage{amsmath}
\usepackage{amssymb}

\title{Challenges in Quantitative Abstractions for\\Collective Adaptive Systems}
\author{Mirco Tribastone
\institute{IMT School for Advanced Studies Lucca, Italy}\\
\email{mirco.tribastone@imtlucca.it}}
\def\titlerunning{Challenges in Quantitative Abstractions}
\def\authorrunning{Mirco Tribastone}
\begin{document}
\maketitle

\begin{abstract}
Like with most large-scale systems, the evaluation of quantitative properties of collective adaptive systems is an important issue that crosscuts all its development stages, from design (in the case of engineered systems) to runtime monitoring and control. Unfortunately it is a difficult problem to tackle in general, due to the typically high computational cost involved in the analysis. This calls for the development of appropriate quantitative abstraction techniques that preserve most of the system's dynamical behaviour using a more compact representation. This paper focuses on models based on ordinary differential equations and reviews recent results where abstraction is achieved by aggregation of variables,   
reflecting on the shortcomings in the state of the art and setting out challenges for future research.
\end{abstract}

\section{Introduction}

Collective adaptive systems (CAS) generally consist of a large number of entities that evolve according to local interactions, across multiple time and space scales, in order to achieve their own objectives. Their behaviour is influenced by such interactions as well as by changes in the environment. As a result, one typically deals with a highly complex system where the overall dynamics cannot be directly inferred from the analysis of the individual entities taken in isolation. This poses a serious problem to our capability of reasoning about CAS effectively. Indeed, for analysis purposes the availability of a model (e.g., an analytical description or discrete-event simulation) greatly helps understand the behaviour under a variety of conditions which would be otherwise difficult, or even impossible, to exercise on the concrete system under consideration. However, the impossibility of studying isolated CAS entities in a model requires the construction of the \emph{product space} of the system that explicitly accounts for all possible interactions between agents. Because of the large scale involved, this clearly incurs the well-known problem of state explosion, affecting both qualitative and quantitative analysis efforts. In the latter case, it is worth to note that state explosion is not merely a computational problem: higher level of detail in the model implies a larger parameter space, which requires substantial effort for measurement, calibration, and validation.  

By \emph{abstraction} we mean a range of techniques where the basic idea is to construct a smaller CAS model, with fewer variables and/or parameters, that can yet preserve most of the dynamics of the original one. Such a more compact representation may be useful, for example, to provide a more intelligible description that ignores original low-level details, or to perform computational analyses more efficiently. Many branches of science and engineering that deal with CAS-like systems have acknowledged the importance of developing abstraction techniques, for instance:
\begin{itemize} 
\item Ecological systems are a prototypical example of (natural) CAS with massive populations of entities interacting with and adapting to the environment. And ecology is a discipline where quantitative abstractions have long been understood (e.g.,~\cite{Iwasa1987287}), motivated by the clear presence of multiple levels of hierarchy---from molecules to whole organisms in their  ecosystem---and a high degree of heterogeneity, e.g., individuals of different age and metabolism, at various locations in space. Here, a large body of work has considered abstractions through \emph{aggregations} of large-scale dynamical models of ecosystems, e.g., based on differential or difference equations. Aggregation is a coarsening of the state space that leads to a self-consistent smaller dynamical system where each ``macro-variable'' appropriately represents a group of ``micro-variables'' in the original model. Aggregations have been developed to ignore, e.g., age differences in population models~\cite{Iwasa1987287} or exploit the separation of time scales typical of ecological systems (see~\cite{AugerSurvey2008} for a survey). 

\item In physics, cellular automata are a basic model to describe complex phenomena such as chaos and fractals, arising from simple interactions between agents with small state spaces. \emph{Coarse graining} methods identify groups of neighbouring cells for which an aggregate dynamics can still describe the overall behaviour of the original model~\cite{PhysRevE.73.026203}.

\item In computational systems biology, a wealth of abstraction techniques have been developed to cope with the combinatorial explosion of the state space of mechanistic dynamical models based on ordinary differential equations (ODEs) for the modelling of protein interaction networks (e.g.~\cite{Borisov2005951,Feret21042009}). Here, available methods consider a \emph{covering} of the state space (where a variable may appear in more than one group)~\cite{Feret21042009}; \emph{quotienting}, induced by a partition of the ODE variables (e.g.,~\cite{Feret2012137}); and aggregations exploiting time-scale separation (e.g.,~\cite{murray:i}). 

\item Large-scale dynamical systems are routinely encountered in control engineering. Here, starting with Aoki~\cite{1098900}, numerous approaches have considered the transformation of the original model into a reduced one that preserves \emph{controllability}, i.e., the capability of bringing the system to a desired state by an appropriate choice of control inputs~\cite{bisimulation_lin_sys_Pappas}.   
\end{itemize}

A common feature shared by these examples is that CAS can be effectively described using dynamical systems. Even when the original description is a Markov chain, the underlying behaviour is characterised by a (large-scale) system of difference/differential equations (i.e., the forward equations of motion of the probability distribution). This paper briefly reviews recently developed techniques that consider the abstraction problem from an algorithmic viewpoint~\cite{concur15,popl16,tacas16}, with the intent of computing reduced dynamical systems that enjoy the following properties:
\begin{enumerate}[\bfseries P1.]
\item The abstraction should come with formal guarantees on the relationship between the abstract dynamics and the original one. This enables the modeller to use the abstract model with full confidence in the results of the analysis.
\item The construction of the abstract model should be fully automatic, since the original model is likely to be unintelligible due to size.
\item The method should be generic in order to be applicable to as wide a range of CAS models as possible.
\item The abstract model should preserve user-defined observables of the original system. For instance, it should be possible to fully recover the dynamics of selected variables of the original model.
\end{enumerate}

The techniques discussed here revolve around the notion of \emph{differential equivalence}, an equivalence relation over the variables of a dynamical system that induces a reduced model where each macro-variable represents the aggregate dynamics of an equivalence class. Although the theory considers first-order ODEs, it straightforwardly carries over to their discrete-time analogues. After a brief introduction in Section~\ref{sec:de}, some challenges for future research in this area are discussed in detail in Section~\ref{sec:outlook}.

\section{Differential Equivalences}\label{sec:de}

According to the terminology above, differential equivalences induce a quotienting of the ODE variables. Two distinct flavours have been provided in~\cite{popl16}.  \emph{Forward differential equivalence} (FDE) is such that a macro-variable describes the sum of the variables of an equivalence class. For instance, consider:
\begin{align}\label{eq:simple}
\dot{x}_1 & = - x_1, &  \dot{x}_2 & = k_1 \cdot x_1 - x_2, & \dot{x}_3 & = k_2 \cdot x_1 - x_3,
\end{align}
where $k_1$ and $k_2$ are constants and the `dot' operator denotes the derivative operator (with respect to time). Then, $\{ \{x_1\}, \{x_2,x_3\}\}$ is an FDE because
\begin{align}\label{eq:simple.fde}
\dot{x}_1 & = - x_1,  & \dot{(x_2 + x_3)} & = \dot{x}_2 + \dot{x}_3 = (k_1 + k_2) \cdot x_1 - (x_2 + x_3) .
\end{align}
By the change of variable $y = x_2 + x_3$, this is equivalent to writing
\begin{align*}
\dot{x}_1 & = - x_1 & \dot{y} & = (k_1 + k_2) \cdot x_1 - y .
\end{align*}
In this quotient model, whenever the \emph{initial condition} (at time point $0$) satisfies $y(0) = x_2(0) + x_3(0)$ we get that $y(t) = x_2(t) + x_3(t)$ at all time points $t$.

\emph{Backward differential equivalence} (BDE) equates variables that have the same solutions at all time points. In (\ref{eq:simple}), $\{ \{x_1\}, \{x_2,x_3\}\}$ is also a BDE provided that $k_1 = k_2$. In this case, we obtain a quotient ODE by removing either equation between $x_2$ and $x_3$, say $x_3$, and rewriting every occurrence of $x_3$ as $x_2$:
\begin{align*}
\dot{x}_1 & = - x_1 & \dot{x}_2 & = k_1 \cdot  x_1 - x_2 .
\end{align*}

Both FDE and BDE satisfy \textbf{P1} because the relationship between the original model and the abstract one is exact; there is, however, loss of information when FDE is applied because the individual traces of the members of an equivalence class may not be recovered in general. Differential equivalences are closely related to the notion of exact ODE lumpability, very well understood in the chemistry literature (e.g.,~\cite{LiRabitz1997,okino1998,Li19891413}). However  this approach lacks of an automatic way of identifying lumping schemes (e.g.,~\cite{Turanyi14}). To cope with this, i.e., to satisfy \textbf{P2}, restrictions are imposed to be able to develop minimisation algorithms.

\paragraph*{Symbolic minimisation.}  In~\cite{popl16} each ODE variable is treated explicitly as a real function and a differential equivalence is encoded in a logical formula over ODE variables. Thus, checking whether a candidate partition is BDE/FDE can be done \emph{symbolically} using an encoding into satisfiability modulo theories (SMT)~\cite{Biere:2009:HSV:1550723}.  In fact, differential equivalences belong to the quantifier-free fragment of first-order logic. It is possible to restrict the admissible ODE systems to those for which an SMT solver for nonlinear real arithmetic --- e.g.,  Z3~\cite{DeMoura:2008:ZES:1792734.1792766} --- is a decision procedure. This can be done by, roughly speaking, excluding trigonometric functions (somewhat satisfying \textbf{P3}). 

Let us consider the example (\ref{eq:simple}), assuming $k_1 = k_2 = 1$. The condition for $\{ \{x_1\}, \{x_2,x_3\}\}$ to be a BDE can be shown to correspond to requiring that related variables with equal assignments always have equal derivatives. This can be encoded in a logical formula $\phi$ thus:
$$ \phi := x_2 = x_3 \Rightarrow k_1 \cdot x_1 - x_2 = k_2 \cdot x_1 - x_3 . $$   The SMT check $\textsl{sat}(\neg \phi)$ looks for an assignment of the variables $x_1$, $x_2$, and $x_3$ for which $\neg \phi$ holds. Thus, the partition is a BDE if and only if the procedure returns ``unsat'' (as it is obviously the case in this example). More interestingly, it is possible to exploit the ability of the solver to return a \emph{witness} in case of satisfiability. 
This can be interpreted as a counterexample that distinguishes variables originally supposed to be equivalent. For instance, an SMT check for the candidate BDE partition $\{ \{ x_1, x_2, x_3 \} \}$ might return the witness $(x_1 = 1, x_2 = 1, x_3 = 1)$, which yields derivatives that are not equivalent $(\dot{x}_1 = -1, \dot{x}_2 = 0, \dot{x}_3 = 0)$. This suggests the implementation of an algorithm that \emph{splits} the partition in such a way to preserve the equalities in the witness. That is, at the next iteration the candidate BDE partition would be  $\{ \{x_1\}, \{x_2,x_3\}\}$. It turns out that such an algorithm does iteratively compute the largest BDE that refines a given initial partition of ODE variables. This algorithm also meets \textbf{P4}: indeed each variable that should be treated as an \emph{observable} can be put in a singleton initial block.

\paragraph*{Syntax-driven minimisation.}  A more efficient minimisation algorithm can be provided for ODEs with derivatives that are multivariate polynomials of degree at most two~\cite{tacas16}. This covers the ubiquitous linear systems as well as chemical reaction networks, at the basis of the aforementioned dynamic models in systems biology. At the basis of this approach is a finitary representation of an ODE system as a so-called \emph{reaction network} (RN), consisting of species/variables interacting by means of reactions parameterised by a real value. On this representation two bisimulation equivalences, the forward and backward RN bisimulations, are related to FDE and BDE, respectively~\cite{concur15}. This makes such bisimulations similar in spirit to quantitative equivalences on labelled transition systems, e.g.,  Larsen and Skou's probabilistic bisimulation~\cite{Larsen19911}. In particular, the computation of the largest RN bisimulations that refine a given partition can be computed using an appropriate variant of Paige and Tarjan's famous algorithm~\cite{partitionref}. In~\cite{tacas16} a partition refinement algorithm is developed along the lines of efficient analogues for Markov chain lumping such as~\cite{Derisavi2003309} and~\cite{DBLP:conf/tacas/ValmariF10}, and for probabilistic transition systems~\cite{DBLP:journals/jcss/BaierEM00}. This computes the largest FB/BB refining a given partition of variables in $O(mn \log n)$ time, where $m$ is the number of monomials in the ODE system and $n$ is the number of variables.

\section{Outlook}\label{sec:outlook}

The differential equivalences reviewed in this paper allow automatic reductions of ODE systems. The benchmarks in~\cite{tacas16} show that the algorithms for reduction up to forward and backward bisimulations can scale to systems with millions of variables and monomials, terminating in a few seconds also in some challenging models. There are, however, further challenges ahead which we wish to tackle.

Forward bisimulation is only a sufficient condition for FDE; while it has been shown to yield significant reductions in practice~\cite{concur15,tacas16}, some other examples from the literature demonstrate that the algorithm may miss some FDE reductions (see~\cite{popl16}). Ongoing work is aiming to develop a new variant of forward bisimulation that characterises FDE (for multivariate polynomial derivatives of degree at most two). Backward bisimulation, on the other hand, does characterise BDE. In this case, the specialised partition-refinement algorithm of~\cite{tacas16} is to be preferred over the symbolic SMT-based approach of \cite{popl16} for computational reasons, since we observed runtimes  generally separated by two/three orders of magnitude in our experiments. This is an unsurprising fact because the SMT-based approach does not exploit the finitary RN representation of~\cite{tacas16}, where no symbolic computation is performed. We plan to exploit the advantage of the RN representation, extending~\cite{tacas16} to polynomial derivatives with arbitrary degree.  

The whole SMT technology has been treated as a black box in our proof-of-concept experiments, despite the well-known fact that appropriate heuristics for SMT can sensibly affect the runtimes. Developing problem-specific solution strategies for differential equivalences is an interesting line of research; a natural question to investigate is the possibility of parallelising the SMT checks at every iteration of the partition refinement. Improving the scalability of the SMT approach will be advantageous for all such models that do not feature degree-two polynomial derivatives. Examples are: third-body reactions (giving degree-three polynomials), appearing frequently in chemical engineering~\cite{Turanyi14}; chemical reaction networks that are based on different kinetics than mass-action, such as Hill's~\cite{Voit:2013aa}; and ODE performance models of computing systems, such as those based on minimum-rate semantics (e.g., stochastic process algebra~\cite{DBLP:journals/tse/TribastoneGH12} and queuing networks~\cite{10.1109/TSE.2012.66}). In addition, also the models that admit the RN representation may benefit from  improvements made in our SMT approach. For instance, the symbolic checks may be used in models with uncertainties in rates (a well-known issue in mathematical biology): here the SMT framework can already be easily extended to compute partitions that are differential equivalences under all possible assignments of such uncertain parameters, left as free variables in the satisfiability problem (similarly to the SMT-based parametric minimisation approach of~\cite{DKP13} for probabilistic models written in PRISM~\cite{DBLP:conf/cav/KwiatkowskaNP11}). 

Lastly, it has long been argued that \emph{exact} reduction techniques such as the ones presented here are too restrictive because they are sensitive to the values of the parameters. It is natural to consider approximate variants (e.g.,~\cite{worrell:approximating,DBLP:journals/tcs/DesharnaisGJP04,DBLP:conf/mfcs/LarsenMP12}) as weaker notions that, for instance, allow variability in the parameters, considering the exact versions as a degenerate case in which no such variability is needed. We remark that, at least for the numerical benchmarks considered so far with realistic models, the exact reductions can already be quite effective. Of course, approximate ones might be able to provide even coarser descriptions. In this case, however, the main challenge is to be able to relate the variability in the parameters tolerated by the coarsening procedure with the error incurred when considering an approximate, smaller model, instead of the original one. This issue has been attempted by some previous work which has provided an asymptotic result of correctness of the approximation for small enough perturbations~\cite{dsn13}. This has been further improved upon using differential inequalities in~\cite{tac15}, providing more usable bounds that however tend to degrade for larger perturbations. In this respect,  two orthogonal lines of research may look into the problem of improving the quality of the bounds and automatically detecting near-symmetries to synthesise candidate approximately reduced models.

\paragraph*{Acknowledgement.} This work is partially supported by the EU project QUANTICOL (\url{http://www.quanticol.eu}), 600708. The author is indebted to Luca Cardelli, Max Tschaikowski, and Andrea Vandin for their collaboration in~\cite{concur15,popl16,tacas16} and the numerous fruitful discussions.
\bibliographystyle{eptcs}
\bibliography{forecast2016}

\begin{thebibliography}{10}
\providecommand{\bibitemdeclare}[2]{}
\providecommand{\surnamestart}{}
\providecommand{\surnameend}{}
\providecommand{\urlprefix}{Available at }
\providecommand{\url}[1]{\texttt{#1}}
\providecommand{\href}[2]{\texttt{#2}}
\providecommand{\urlalt}[2]{\href{#1}{#2}}
\providecommand{\doi}[1]{doi:\urlalt{http://dx.doi.org/#1}{#1}}
\providecommand{\bibinfo}[2]{#2}

\bibitemdeclare{article}{1098900}
\bibitem{1098900}
\bibinfo{author}{M.~\surnamestart Aoki\surnameend} (\bibinfo{year}{1968}):
  \emph{\bibinfo{title}{Control of large-scale dynamic systems by
  aggregation}}.
\newblock {\sl \bibinfo{journal}{IEEE Trans. Autom. Control}}
  \bibinfo{volume}{13}(\bibinfo{number}{3}), pp. \bibinfo{pages}{246--253},
  \doi{10.1109/TAC.1968.1098900}.

\bibitemdeclare{incollection}{AugerSurvey2008}
\bibitem{AugerSurvey2008}
\bibinfo{author}{P.~\surnamestart Auger\surnameend},
  \bibinfo{author}{R.~\surnamestart de~la Parra\surnameend},
  \bibinfo{author}{J.~\surnamestart Poggiale\surnameend},
  \bibinfo{author}{E.~\surnamestart S{\'a}nchez\surnameend} \&
  \bibinfo{author}{T.~\surnamestart Nguyen-Huu\surnameend}
  (\bibinfo{year}{2008}): \emph{\bibinfo{title}{Aggregation of Variables and
  Applications to Population Dynamics}}.
\newblock In \bibinfo{editor}{Pierre \surnamestart Magal\surnameend} \&
  \bibinfo{editor}{Shigui \surnamestart Ruan\surnameend}, editors: {\sl
  \bibinfo{booktitle}{Structured Population Models in Biology and
  Epidemiology}}, {\sl \bibinfo{series}{Lecture Notes in Mathematics}}
  \bibinfo{volume}{1936}, \bibinfo{publisher}{Springer}, pp.
  \bibinfo{pages}{209--263}, \doi{10.1007/978-3-540-78273-5\_5}.

\bibitemdeclare{article}{DBLP:journals/jcss/BaierEM00}
\bibitem{DBLP:journals/jcss/BaierEM00}
\bibinfo{author}{C.~\surnamestart Baier\surnameend},
  \bibinfo{author}{B.~\surnamestart Engelen\surnameend} \&
  \bibinfo{author}{M.~E. \surnamestart Majster-Cederbaum\surnameend}
  (\bibinfo{year}{2000}): \emph{\bibinfo{title}{Deciding Bisimilarity and
  Similarity for Probabilistic Processes}}.
\newblock {\sl \bibinfo{journal}{J. Comput. Syst. Sci.}}
  \bibinfo{volume}{60}(\bibinfo{number}{1}), pp. \bibinfo{pages}{187--231},
  \doi{10.1006/jcss.1999.1683}.

\bibitemdeclare{inbook}{Biere:2009:HSV:1550723}
\bibitem{Biere:2009:HSV:1550723}
\bibinfo{author}{C.W. \surnamestart Barrett\surnameend},
  \bibinfo{author}{R.~\surnamestart Sebastiani\surnameend},
  \bibinfo{author}{S.A. \surnamestart Seshia\surnameend} \&
  \bibinfo{author}{C.~\surnamestart Tinelli\surnameend} (\bibinfo{year}{2009}):
  \emph{\bibinfo{title}{Handbook of Satisfiability: Volume 185 Frontiers in
  Artificial Intelligence and Applications}}, chapter
  \bibinfo{chapter}{Satisfiability Modulo Theories}.
\newblock \bibinfo{publisher}{IOS Press}.

\bibitemdeclare{article}{Borisov2005951}
\bibitem{Borisov2005951}
\bibinfo{author}{N.M. \surnamestart Borisov\surnameend}, \bibinfo{author}{N.I.
  \surnamestart Markevich\surnameend}, \bibinfo{author}{J.B. \surnamestart
  Hoek\surnameend} \& \bibinfo{author}{B.N. \surnamestart
  Kholodenko\surnameend} (\bibinfo{year}{2005}):
  \emph{\bibinfo{title}{Signaling through Receptors and Scaffolds: Independent
  Interactions Reduce Combinatorial Complexity}}.
\newblock {\sl \bibinfo{journal}{Biophysical Journal}}
  \bibinfo{volume}{89}(\bibinfo{number}{2}), pp. \bibinfo{pages}{951--966},
  \doi{10.1529/biophysj.105.060533}.

\bibitemdeclare{article}{worrell:approximating}
\bibitem{worrell:approximating}
\bibinfo{author}{F.~\surnamestart van Breugel\surnameend} \&
  \bibinfo{author}{J.~\surnamestart Worrell\surnameend} (\bibinfo{year}{2006}):
  \emph{\bibinfo{title}{Approximating and computing behavioural distances in
  probabilistic transition systems}}.
\newblock {\sl \bibinfo{journal}{Theoretical Computer Science}}
  \bibinfo{volume}{360}(\bibinfo{number}{1--3}), pp. \bibinfo{pages}{373--385},
  \doi{10.1016/j.tcs.2006.05.021}.

\bibitemdeclare{inproceedings}{concur15}
\bibitem{concur15}
\bibinfo{author}{L.~\surnamestart Cardelli\surnameend},
  \bibinfo{author}{M.~\surnamestart Tribastone\surnameend},
  \bibinfo{author}{M.~\surnamestart Tschaikowski\surnameend} \&
  \bibinfo{author}{A.~\surnamestart Vandin\surnameend} (\bibinfo{year}{2015}):
  \emph{\bibinfo{title}{Forward and Backward Bisimulations for Chemical
  Reaction Networks}}.
\newblock In: {\sl \bibinfo{booktitle}{26th International Conference on
  Concurrency Theory, {CONCUR}}}, pp. \bibinfo{pages}{226--239},
  \doi{10.4230/LIPIcs.CONCUR.2015.226}.

\bibitemdeclare{inproceedings}{tacas16}
\bibitem{tacas16}
\bibinfo{author}{L.~\surnamestart Cardelli\surnameend},
  \bibinfo{author}{M.~\surnamestart Tribastone\surnameend},
  \bibinfo{author}{M.~\surnamestart Tschaikowski\surnameend} \&
  \bibinfo{author}{A.~\surnamestart Vandin\surnameend} (\bibinfo{year}{2016}):
  \emph{\bibinfo{title}{Efficient Syntax-driven Lumping of Differential
  Equations}}.
\newblock In: {\sl \bibinfo{booktitle}{Proceedings of Tools and Algorithms for
  the Construction and Analysis of Systems --- 21st International Conference
  (TACAS)}}, \doi{10.1007/978-3-662-49674-9\_6}.

\bibitemdeclare{inproceedings}{popl16}
\bibitem{popl16}
\bibinfo{author}{L.~\surnamestart Cardelli\surnameend},
  \bibinfo{author}{M.~\surnamestart Tribastone\surnameend},
  \bibinfo{author}{M.~\surnamestart Tschaikowski\surnameend} \&
  \bibinfo{author}{A.~\surnamestart Vandin\surnameend} (\bibinfo{year}{2016}):
  \emph{\bibinfo{title}{Symbolic computation of differential equivalences}}.
\newblock In: {\sl \bibinfo{booktitle}{Proceedings of the 43rd Annual {ACM}
  {SIGPLAN-SIGACT} Symposium on Principles of Programming Languages, {POPL}}},
  pp. \bibinfo{pages}{137--150}, \doi{10.1145/2837614.2837649}.

\bibitemdeclare{inproceedings}{DeMoura:2008:ZES:1792734.1792766}
\bibitem{DeMoura:2008:ZES:1792734.1792766}
\bibinfo{author}{L.~\surnamestart De~Moura\surnameend} \&
  \bibinfo{author}{N.~\surnamestart Bj\o{}rner\surnameend}
  (\bibinfo{year}{2008}): \emph{\bibinfo{title}{{Z3}: An Efficient {SMT}
  Solver}}.
\newblock In: {\sl \bibinfo{booktitle}{TACAS}}, pp. \bibinfo{pages}{337--340},
  \doi{10.1007/978-3-540-78800-3\_24}.

\bibitemdeclare{inproceedings}{DKP13}
\bibitem{DKP13}
\bibinfo{author}{C.~\surnamestart Dehnert\surnameend}, \bibinfo{author}{J.-P.
  \surnamestart Katoen\surnameend} \& \bibinfo{author}{D.~\surnamestart
  Parker\surnameend} (\bibinfo{year}{2013}): \emph{\bibinfo{title}{{SMT}-Based
  Bisimulation Minimisation of {Markov} Models}}.
\newblock In \bibinfo{editor}{R.~\surnamestart Giacobazzi\surnameend},
  \bibinfo{editor}{J.~\surnamestart Berdine\surnameend} \&
  \bibinfo{editor}{I.~\surnamestart Mastroeni\surnameend}, editors: {\sl
  \bibinfo{booktitle}{Proc. 14th International Conference on Verification,
  Model Checking, and Abstract Interpretation (VMCAI'13)}}, {\sl
  \bibinfo{series}{LNCS}} \bibinfo{volume}{7737},
  \bibinfo{publisher}{Springer}, pp. \bibinfo{pages}{28--47},
  \doi{10.1007/978-3-642-35873-9\_5}.

\bibitemdeclare{article}{Derisavi2003309}
\bibitem{Derisavi2003309}
\bibinfo{author}{S.~\surnamestart Derisavi\surnameend},
  \bibinfo{author}{H.~\surnamestart Hermanns\surnameend} \&
  \bibinfo{author}{W.H. \surnamestart Sanders\surnameend}
  (\bibinfo{year}{2003}): \emph{\bibinfo{title}{Optimal state-space lumping in
  {M}arkov chains}}.
\newblock {\sl \bibinfo{journal}{Information Processing Letters}}
  \bibinfo{volume}{87}(\bibinfo{number}{6}), pp. \bibinfo{pages}{309--315},
  \doi{10.1016/S0020-0190(03)00343-0}.

\bibitemdeclare{article}{DBLP:journals/tcs/DesharnaisGJP04}
\bibitem{DBLP:journals/tcs/DesharnaisGJP04}
\bibinfo{author}{J.~\surnamestart Desharnais\surnameend},
  \bibinfo{author}{V.~\surnamestart Gupta\surnameend},
  \bibinfo{author}{R.~\surnamestart Jagadeesan\surnameend} \&
  \bibinfo{author}{P.~\surnamestart Panangaden\surnameend}
  (\bibinfo{year}{2004}): \emph{\bibinfo{title}{Metrics for labelled Markov
  processes}}.
\newblock {\sl \bibinfo{journal}{Theor. Comput. Sci.}}
  \bibinfo{volume}{318}(\bibinfo{number}{3}), pp. \bibinfo{pages}{323--354},
  \doi{10.1016/j.tcs.2003.09.013}.

\bibitemdeclare{article}{Feret21042009}
\bibitem{Feret21042009}
\bibinfo{author}{J.~\surnamestart Feret\surnameend},
  \bibinfo{author}{V.~\surnamestart Danos\surnameend},
  \bibinfo{author}{J.~\surnamestart Krivine\surnameend},
  \bibinfo{author}{R.~\surnamestart Harmer\surnameend} \&
  \bibinfo{author}{W.~\surnamestart Fontana\surnameend} (\bibinfo{year}{2009}):
  \emph{\bibinfo{title}{Internal coarse-graining of molecular systems}}.
\newblock {\sl \bibinfo{journal}{Proceedings of the National Academy of
  Sciences}} \bibinfo{volume}{106}(\bibinfo{number}{16}), pp.
  \bibinfo{pages}{6453--6458}, \doi{10.1073/pnas.0809908106}.

\bibitemdeclare{article}{Feret2012137}
\bibitem{Feret2012137}
\bibinfo{author}{J.~\surnamestart Feret\surnameend},
  \bibinfo{author}{T.~\surnamestart Henzinger\surnameend},
  \bibinfo{author}{H.~\surnamestart Koeppl\surnameend} \&
  \bibinfo{author}{T.~\surnamestart Petrov\surnameend} (\bibinfo{year}{2012}):
  \emph{\bibinfo{title}{Lumpability abstractions of rule-based systems}}.
\newblock {\sl \bibinfo{journal}{Theoretical Computer Science}}
  \bibinfo{volume}{431}, pp. \bibinfo{pages}{137--164},
  \doi{10.1016/j.tcs.2011.12.059}.

\bibitemdeclare{inproceedings}{dsn13}
\bibitem{dsn13}
\bibinfo{author}{G.~\surnamestart Iacobelli\surnameend} \&
  \bibinfo{author}{M.~\surnamestart Tribastone\surnameend}
  (\bibinfo{year}{2013}): \emph{\bibinfo{title}{Lumpability of fluid models
  with heterogeneous agent types}}.
\newblock In: {\sl \bibinfo{booktitle}{DSN}}, pp. \bibinfo{pages}{1--11},
  \doi{10.1109/DSN.2013.6575346}.

\bibitemdeclare{article}{PhysRevE.73.026203}
\bibitem{PhysRevE.73.026203}
\bibinfo{author}{N.~\surnamestart Israeli\surnameend} \&
  \bibinfo{author}{N.~\surnamestart Goldenfeld\surnameend}
  (\bibinfo{year}{2006}): \emph{\bibinfo{title}{Coarse-graining of cellular
  automata, emergence, and the predictability of complex systems}}.
\newblock {\sl \bibinfo{journal}{Phys. Rev. E}} \bibinfo{volume}{73}, p.
  \bibinfo{pages}{026203}, \doi{10.1103/PhysRevE.73.026203}.

\bibitemdeclare{article}{Iwasa1987287}
\bibitem{Iwasa1987287}
\bibinfo{author}{Y.~\surnamestart Iwasa\surnameend},
  \bibinfo{author}{V.~\surnamestart Andreasen\surnameend} \&
  \bibinfo{author}{S.~\surnamestart Levin\surnameend} (\bibinfo{year}{1987}):
  \emph{\bibinfo{title}{{Aggregation in model ecosystems. I. Perfect
  aggregation}}}.
\newblock {\sl \bibinfo{journal}{Ecological Modelling}}
  \bibinfo{volume}{37}(\bibinfo{number}{3-4}), pp. \bibinfo{pages}{287--302},
  \doi{10.1016/0304-3800(87)90030-5}.

\bibitemdeclare{inproceedings}{DBLP:conf/cav/KwiatkowskaNP11}
\bibitem{DBLP:conf/cav/KwiatkowskaNP11}
\bibinfo{author}{M.Z. \surnamestart Kwiatkowska\surnameend},
  \bibinfo{author}{G.~\surnamestart Norman\surnameend} \&
  \bibinfo{author}{D.~\surnamestart Parker\surnameend} (\bibinfo{year}{2011}):
  \emph{\bibinfo{title}{PRISM 4.0: Verification of Probabilistic Real-Time
  Systems}}.
\newblock In: {\sl \bibinfo{booktitle}{CAV}}, pp. \bibinfo{pages}{585--591},
  \doi{10.1007/978-3-642-22110-1\_47}.

\bibitemdeclare{inproceedings}{DBLP:conf/mfcs/LarsenMP12}
\bibitem{DBLP:conf/mfcs/LarsenMP12}
\bibinfo{author}{K.G. \surnamestart Larsen\surnameend},
  \bibinfo{author}{R.~\surnamestart Mardare\surnameend} \&
  \bibinfo{author}{P.~\surnamestart Panangaden\surnameend}
  (\bibinfo{year}{2012}): \emph{\bibinfo{title}{Taking It to the Limit:
  Approximate Reasoning for Markov Processes}}.
\newblock In: {\sl \bibinfo{booktitle}{MFCS}}, pp. \bibinfo{pages}{681--692},
  \doi{10.1007/978-3-642-32589-2\_59}.

\bibitemdeclare{article}{Larsen19911}
\bibitem{Larsen19911}
\bibinfo{author}{K.G. \surnamestart Larsen\surnameend} \&
  \bibinfo{author}{A.~\surnamestart Skou\surnameend} (\bibinfo{year}{1991}):
  \emph{\bibinfo{title}{Bisimulation through probabilistic testing}}.
\newblock {\sl \bibinfo{journal}{Information and Computation}}
  \bibinfo{volume}{94}(\bibinfo{number}{1}), pp. \bibinfo{pages}{1--28},
  \doi{10.1016/0890-5401(91)90030-6}.

\bibitemdeclare{article}{Li19891413}
\bibitem{Li19891413}
\bibinfo{author}{G.~\surnamestart Li\surnameend} \&
  \bibinfo{author}{H.~\surnamestart Rabitz\surnameend} (\bibinfo{year}{1989}):
  \emph{\bibinfo{title}{A general analysis of exact lumping in chemical
  kinetics}}.
\newblock {\sl \bibinfo{journal}{Chemical Engineering Science}}
  \bibinfo{volume}{44}(\bibinfo{number}{6}), pp. \bibinfo{pages}{1413--1430},
  \doi{10.1016/0009-2509(89)85014-6}.

\bibitemdeclare{book}{murray:i}
\bibitem{murray:i}
\bibinfo{author}{J.D. \surnamestart Murray\surnameend} (\bibinfo{year}{2002}):
  \emph{\bibinfo{title}{Mathematical Biology I: An Introduction}},
  \bibinfo{edition}{3rd} edition.
\newblock \bibinfo{publisher}{Springer}, \doi{10.1007/b98868}.

\bibitemdeclare{article}{okino1998}
\bibitem{okino1998}
\bibinfo{author}{M.S. \surnamestart Okino\surnameend} \& \bibinfo{author}{M.L.
  \surnamestart Mavrovouniotis\surnameend} (\bibinfo{year}{1998}):
  \emph{\bibinfo{title}{Simplification of Mathematical Models of Chemical
  Reaction Systems}}.
\newblock {\sl \bibinfo{journal}{Chemical Reviews}}
  \bibinfo{volume}{2}(\bibinfo{number}{98}), pp. \bibinfo{pages}{391--408},
  \doi{10.1021/cr950223l}.

\bibitemdeclare{article}{partitionref}
\bibitem{partitionref}
\bibinfo{author}{R.~\surnamestart Paige\surnameend} \&
  \bibinfo{author}{R.~\surnamestart Tarjan\surnameend} (\bibinfo{year}{1987}):
  \emph{\bibinfo{title}{Three Partition Refinement Algorithms}}.
\newblock {\sl \bibinfo{journal}{SIAM Journal on Computing}}
  \bibinfo{volume}{16}(\bibinfo{number}{6}), pp. \bibinfo{pages}{973--989},
  \doi{10.1137/0216062}.

\bibitemdeclare{article}{bisimulation_lin_sys_Pappas}
\bibitem{bisimulation_lin_sys_Pappas}
\bibinfo{author}{G.J. \surnamestart Pappas\surnameend} (\bibinfo{year}{2003}):
  \emph{\bibinfo{title}{Bisimilar linear systems}}.
\newblock {\sl \bibinfo{journal}{Automatica}}
  \bibinfo{volume}{39}(\bibinfo{number}{12}), pp. \bibinfo{pages}{2035--2047},
  \doi{10.1016/j.automatica.2003.07.003}.

\bibitemdeclare{article}{LiRabitz1997}
\bibitem{LiRabitz1997}
\bibinfo{author}{J.~\surnamestart Toth\surnameend},
  \bibinfo{author}{G.~\surnamestart Li\surnameend},
  \bibinfo{author}{H.~\surnamestart Rabitz\surnameend} \& \bibinfo{author}{A.S.
  \surnamestart Tomlin\surnameend} (\bibinfo{year}{1997}):
  \emph{\bibinfo{title}{The Effect of Lumping and Expanding on Kinetic
  Differential Equations}}.
\newblock {\sl \bibinfo{journal}{SIAM Journal on Applied Mathematics}}
  \bibinfo{volume}{57}(\bibinfo{number}{6}), pp. \bibinfo{pages}{1531--1556},
  \doi{10.1137/S0036139995293294}.

\bibitemdeclare{article}{10.1109/TSE.2012.66}
\bibitem{10.1109/TSE.2012.66}
\bibinfo{author}{M.~\surnamestart Tribastone\surnameend}
  (\bibinfo{year}{2013}): \emph{\bibinfo{title}{A Fluid Model for Layered
  Queueing Networks}}.
\newblock {\sl \bibinfo{journal}{IEEE Transactions on Software Engineering}}
  \bibinfo{volume}{39}(\bibinfo{number}{6}), pp. \bibinfo{pages}{744--756},
  \doi{10.1109/TSE.2012.66}.

\bibitemdeclare{article}{DBLP:journals/tse/TribastoneGH12}
\bibitem{DBLP:journals/tse/TribastoneGH12}
\bibinfo{author}{M.~\surnamestart Tribastone\surnameend},
  \bibinfo{author}{S.~\surnamestart Gilmore\surnameend} \&
  \bibinfo{author}{J.~\surnamestart Hillston\surnameend}
  (\bibinfo{year}{2012}): \emph{\bibinfo{title}{Scalable Differential Analysis
  of Process Algebra Models}}.
\newblock {\sl \bibinfo{journal}{IEEE Transactions on Software Engineering}}
  \bibinfo{volume}{38}(\bibinfo{number}{1}), pp. \bibinfo{pages}{205--219},
  \doi{10.1109/TSE.2010.82}.

\bibitemdeclare{article}{tac15}
\bibitem{tac15}
\bibinfo{author}{M.~\surnamestart Tschaikowski\surnameend} \&
  \bibinfo{author}{M.~\surnamestart Tribastone\surnameend}
  (\bibinfo{year}{2015}): \emph{\bibinfo{title}{Approximate reduction of
  heterogeneous nonlinear models with differential hulls}}.
\newblock {\sl \bibinfo{journal}{IEEE Transactions on Automatic Control}},
  \doi{10.1109/TAC.2015.2457172}.

\bibitemdeclare{book}{Turanyi14}
\bibitem{Turanyi14}
\bibinfo{author}{T.~\surnamestart Turanyi\surnameend} \&
  \bibinfo{author}{A.~\surnamestart Tomlin\surnameend} (\bibinfo{year}{2014}):
  \emph{\bibinfo{title}{Analysis of Kinetic Reaction Mechanisms}}.
\newblock \bibinfo{publisher}{Springer}, \doi{10.1007/978-3-662-44562-4}.

\bibitemdeclare{inproceedings}{DBLP:conf/tacas/ValmariF10}
\bibitem{DBLP:conf/tacas/ValmariF10}
\bibinfo{author}{A.~\surnamestart Valmari\surnameend} \&
  \bibinfo{author}{G.~\surnamestart Franceschinis\surnameend}
  (\bibinfo{year}{2010}): \emph{\bibinfo{title}{Simple \emph{O}(\emph{m}
  log\emph{n}) Time Markov Chain Lumping}}.
\newblock In: {\sl \bibinfo{booktitle}{TACAS}}, pp. \bibinfo{pages}{38--52},
  \doi{10.1007/978-3-642-12002-2\_4}.

\bibitemdeclare{article}{Voit:2013aa}
\bibitem{Voit:2013aa}
\bibinfo{author}{Eberhard~O. \surnamestart Voit\surnameend}
  (\bibinfo{year}{2013}): \emph{\bibinfo{title}{Biochemical Systems Theory: A
  Review}}.
\newblock {\sl \bibinfo{journal}{ISRN Biomathematics}} \bibinfo{volume}{2013},
  p.~\bibinfo{pages}{53}, \doi{10.1155/2013/897658}.

\end{thebibliography}
\end{document}